\documentclass[12pt,showpacs,showkeys,amsmath,amssymb]{revtex4}
\usepackage{amsmath,amsfonts,amsthm,amscd,amssymb,latexsym}
\usepackage{bm}
\usepackage{dcolumn}
\usepackage{graphicx}
\usepackage{epstopdf}
\usepackage{color}
\usepackage{epsf}
\usepackage{epsfig}
\usepackage{graphicx, epic, eepic, color}

\newcommand{\beq}{\begin{equation}}
\newcommand{\eeq}{\end{equation}}

\def\a{\alpha}
\def\b{\beta}
\def\g{\gamma}

\def\d{\delta}

\def\@{\partial_}

\def\negenspace{\kern-1.1em}

\def\sqr#1#2{{\vcenter{\hrule height.#2pt\hbox{\vrule width.#2pt
height#1pt \kern#1pt \vrule width.#2pt}\hrule height.#2pt}}}
\def\square{\mathchoice\sqr64\sqr64\sqr{4.2}3\sqr{3.0}3}

\def\bfx{\mathbf{x}}

\def\stareq{\stackrel{*}{=}}

\begin{document}

\title{Nonlocal Gravitomagnetism}

\author{Bahram \surname{Mashhoon}$^{1,2}$}
\email{mashhoonb@missouri.edu}
\author{Friedrich W. \surname{Hehl}$^{3}$}
\email{hehl@thp.uni-koeln.de}

\affiliation{$^1$Department of Physics and Astronomy, University of Missouri, Columbia, Missouri 65211, USA\\
$^2$School of Astronomy, Institute for Research in Fundamental
Sciences (IPM), P. O. Box 19395-5531, Tehran, Iran\\
$^3$Institute for Theoretical Physics, University of Cologne, 50923 K\"oln, Germany\\
}

\date{\today}

\begin{abstract}
We briefly review the current status of nonlocal gravity (NLG), which is a classical nonlocal generalization of Einstein's theory of gravitation based on a certain analogy with the nonlocal electrodynamics of media. Nonlocal gravity thus involves integro-differential field equations and a causal constitutive kernel that should ultimately be determined from observational data. We consider the stationary gravitational field of an isolated rotating astronomical source in the linear approximation of nonlocal gravity. In this weak-field and slow-motion approximation of NLG, we describe the gravitomagnetic field associated with the rotating source and compare our results with gravitoelectromagnetism (GEM) of the standard general relativity theory. Moreover, we briefly  study the energy-momentum content of the GEM field in nonlocal gravity. 
\end{abstract}


\maketitle

\section{Introduction}

The standard formulation of general relativity (GR) involves the extension of classical physics expressed in Minkowski spacetime, with metric $dS^2 = \eta_{\mu \nu}\,dX^\mu \,dX^\nu$, first to arbitrary curvilinear (``accelerated") coordinates via the locality postulate and then to curved spacetime, with metric $ds^2 = g_{\mu \nu} \,dx^\mu \,dx^\nu$, by means of Einstein's principle of equivalence~\cite{Einstein, Mash1, Mash2}. Here,   $\eta_{\alpha \beta}$ is the Minkowski metric tensor given by diag$(-1,1,1,1)$, latin indices run from 1 to 3, while greek indices run from 0 to 3. The theory is thus based on the Levi-Civita connection
\begin{equation}\label{I1}
{^0}\Gamma^\mu_{\alpha \beta}= \frac{1}{2} g^{\mu \nu} (g_{\nu \alpha,\beta}+g_{\nu \beta,\alpha}-g_{\alpha \beta,\nu})\,.
\end{equation}
This symmetric connection is torsion free, but has Riemannian curvature
\begin{equation}\label{I2}
^{0}R^{\alpha}{}_{\mu \beta \nu}=\partial_{\beta}\, {^0} \Gamma^\alpha_{\nu \mu} -\partial_{\nu}\, {^0} \Gamma^\alpha_{\beta \mu}+\,^{0}\Gamma^{\alpha}_{\beta \gamma}\, ^{0}\Gamma^{\gamma}_{\nu \mu}-\,^{0}\Gamma^\alpha_{\nu \gamma}\, ^{0}\Gamma^\gamma_{\beta \mu}\,.
\end{equation}
A left superscript ``0" will be employed throughout to designate all geometric quantities that are related to the Levi-Civita connection.

In the curved spacetime of general relativity, free test particles and light rays follow timelike and null geodesics, respectively. The correspondence with Newtonian gravitation is established via Einstein's field equations~\cite{Einstein}
\begin{equation}\label{I3}
{^0}G_{\mu \nu} + \Lambda\, g_{\mu \nu}=\kappa\,T_{\mu \nu}\,, \qquad {^0}G_{\mu \nu} := {^0}R_{\mu \nu}-\frac{1}{2} g_{\mu \nu}\,{^0}R\,,
 \end{equation}
where ${^0}G_{\mu \nu}$ is the Einstein tensor, ${^0}R_{\mu \nu}={^0}R^{\alpha}{}_{\mu \alpha \nu}$ is the Ricci tensor and ${^0}R= g^{\mu \nu}\,{^0}R_{\mu \nu}$ is the scalar curvature. Moreover, $T_{\mu \nu}$ is the symmetric energy-momentum tensor of matter (and nongravitational fields), $\Lambda$ is the cosmological constant and $\kappa:=8 \pi G/c^4$.  In GR, the gravitational field is identified with the Riemannian curvature of spacetime; therefore, spacetime is flat when gravity is turned off and we then work within the framework of the special theory of relativity. 

Einstein's general relativity has significant observational support. Indeed,  GR is at present in good agreement with solar system data as well as data from astronomical binary systems. The recent detection of gravitation radiation due to binary mergers lends further support to Einstein's theory of gravitation. On the other hand, in the current standard model of cosmology, which assumes the spatial homogeneity and isotropy of the universe, the energy content of the universe consists of about 70\% dark energy, about 25\% dark matter and about 5\% visible matter. Dark energy is a kind of repulsive energy that permeates the universe and not only counteracts the attraction of matter, but causes accelerated expansion of the universe.  The nature and origin of dark energy are unknown, but it should have positive energy density and negative pressure. It is uniformly distributed throughout space and though it exists everywhere, it is extremely difficult to detect locally. A possible candidate for dark energy is provided by the cosmological constant $\Lambda$. The existence of dark energy and dark matter indicates that we are almost completely ignorant about our universe. Most of the matter in the universe is currently thought to be in the form of certain elusive particles of dark matter that, despite much effort, have not been directly detected. The existence and properties of this dark matter have thus far been deduced only through its gravity. In modern astronomy, dark matter is needed to explain dynamics of galaxies, clusters of galaxies and structure formation in cosmology. However, it is possible that there is no dark matter at all and the theory of gravitation needs to be modified on the scale of galaxies and beyond in order to take due account of what appears as dark matter in astronomy and cosmology. A suitably extended theory of gravitation could then account for the observational data without any need for dark matter. The present paper is about an attempt in this direction; that is, the nonlocal aspect of gravity in NLG simulates dark matter. The main purpose of this paper is to briefly present the main features of NLG theory and develop a useful linear perturbation scheme involving nonlocal gravitoelectromagnetism. 
 
Einstein's theory of gravitation can be alternatively formulated within the framework of teleparallelism. In this approach to gravitation, the fundamental fields are the sixteen components of an arbitrary smooth orthonormal tetrad frame  $e^\mu{}_{\hat{\alpha}}(x)$. The spacetime metric is then defined via the orthonormality condition
\begin{equation}\label{I4}
g_{\mu \nu} = \eta_{\hat{\alpha} \hat{\beta}}\,e_\mu{}^{\hat{\alpha}}(x)\, e_\nu{}^{\hat{\beta}}(x)\,.
\end{equation}
Here, the \emph{hatted} indices (e.g., $\hat {\alpha}$) refer to \emph{anholonomic} tetrad---that is, local Lorentz---indices, while ordinary indices (e.g., $\alpha$) refer to \emph{holonomic} spacetime indices. For instance, in 
\begin{equation}\label{I5}
ds^2 = g_{\mu \nu}\,dx^\mu \, dx^\nu = \eta_{\hat {\alpha} \hat {\beta}}\,dx^{\hat {\alpha}} \, dx^{\hat {\beta}}\,, \quad dx^\mu= e^\mu{}_{\hat {\alpha}}\,dx^{\hat {\alpha}}\,,\quad dx^{\hat {\alpha}} = e_\mu{}^{\hat {\alpha}}\,dx^\mu\,,
\end{equation} 
the tetrad connects (holonomic) spacetime quantities to (anholonomic) local Lorentz quantities. A coordinate basis is holonomic, while a noncoordinate basis is anholonomic. For instance, given a coordinate system $x^\mu$, four coordinate lines pass through each event and for each $\mu = 0, 1, 2, 3$, the 1-form $dx^\mu$ is exact and hence integrable. On the other hand, for each $\hat \alpha = \hat 0, \hat 1, \hat 2, \hat 3$, the 1-form $dx^{\hat {\alpha}}$ in Eq.~\eqref{I5} is in general not exact and hence nonintegrable. Holonomic systems are integrable, while anholonomic systems are nonintegrable. Holonomic and anholonomic indices are raised and lowered by means of $g_{\mu \nu}(x)$ and $\eta_{\hat {\alpha} \hat {\beta}}$, respectively. To change an anholonomic index of a tensor into a holonomic index or vice versa, we simply project the tensor onto the corresponding tetrad frame. We use units such that $c=1$, unless specified otherwise.

The chosen tetrad frame is employed to define the \emph{Weitzenb\"ock connection}~\cite{We}
\begin{equation}\label{I6}
\Gamma^\mu_{\alpha \beta}=e^\mu{}_{\hat{\rho}}~\partial_\alpha\,e_\beta{}^{\hat{\rho}}\,.
\end{equation}
This nonsymmetric connection is curvature free, but has torsion. It follows from definition~\eqref{I6} that the tetrad frame is covariantly constant
\begin{equation}\label{I7}
\nabla_\nu\,e_\mu{}^{\hat{\alpha}}=0\,,
\end{equation}
where $\nabla$ refers to covariant differentiation with respect to the Weitzenb\"ock connection.  Equation~\eqref{I7} implies that each leg of the tetrad  field is parallel to itself throughout the manifold. That is, for  each $\hat \alpha$, Eq.~\eqref{I7}  is an expression of the parallel  transport of the corresponding vector with respect to connection~\eqref{I6}.
Thus in this theory observers throughout spacetime have access to a global set of parallel vector fields that constitute the components of the tetrad frame field. This circumstance is the essence of teleparallelism; for example, two distant vectors can be considered parallel to each other if they have the same components with respect to the local tetrad frames.

 It follows from Eqs.~\eqref{I4} and~\eqref{I7} that $\nabla_\gamma\, g_{\alpha \beta}=0$, so that the Weitzenb\"ock connection is compatible with the metric. Thus in the framework under consideration here, we have one spacetime metric and two metric-compatible connections. It is therefore possible to introduce the \emph{torsion} tensor
\begin{equation}\label{I8}
 C_{\mu \nu}{}^{\alpha}=\Gamma^{\alpha}_{\mu \nu}-\Gamma^{\alpha}_{\nu \mu}=e^\alpha{}_{\hat{\beta}}\Big(\partial_{\mu}e_{\nu}{}^{\hat{\beta}}-\partial_{\nu}e_{\mu}{}^{\hat{\beta}}\Big)\,,
\end{equation}
and the \emph{contorsion} tensor
\begin{equation}\label{I9}
K_{\mu \nu}{}^\alpha= {^0} \Gamma^\alpha_{\mu \nu} - \Gamma^\alpha_{\mu \nu}\,,
\end{equation}
which are linearly related. To see this, we note that $\nabla_\gamma\,g_{\alpha \beta}=0$ implies
\begin{equation}\label{I10}
 g_{\alpha \beta , \gamma}= \Gamma^\mu_{\gamma \alpha}\, g_{\mu \beta} + \Gamma^\mu_{\gamma \beta}\, g_{\mu \alpha}\,,
\end{equation}
which, via the Levi-Civita connection~\eqref{I1}, leads to
\begin{equation}\label{I11}
K_{\mu \nu}{}^\alpha = \frac{1}{2}\, g^{\alpha \beta} (C_{\mu \beta \nu}+C_{\nu \beta \mu}-C_{\mu \nu \beta})\,.
\end{equation}

The torsion tensor is antisymmetric in its first two indices by definition; however, the contorsion tensor turns out to be antisymmetric in its last two indices. The torsion of the Weitzenb\"ock connection and the curvature of the Levi-Civita connection are complementary aspects of the gravitational field within the framework of teleparallelism. Thus it is natural to express Einstein's field equations in terms of the torsion tensor. The result is the teleparallel equivalent of general relativity, GR$_{||}$, to which we now turn. 

\subsection{GR$_{||}$}

It follows from Eqs.~\eqref{I9} and~\eqref{I11} that one can write Einstein's field equations in terms of the torsion tensor. To this end, one can prove after much algebra that the Einstein tensor is given by
\begin{eqnarray}\label{I12}
 {^0}G_{\mu \nu}=\frac{\kappa}{\sqrt{-g}}\Big[e_\mu{}^{\hat{\gamma}}\,g_{\nu \alpha}\, \frac{\partial}{\partial x^\beta}\,{\cal H}^{\alpha \beta}{}_{\hat{\gamma}}
-\Big(C_{\mu}{}^{\rho \sigma}\,{\cal H}_{\nu \rho \sigma}
-\frac{1}{4}\,g_{\mu \nu}\,C^{\alpha \beta \gamma}\,{\cal H}_{\alpha \beta \gamma}\Big) \Big]\,,
\end{eqnarray}
where we have introduced auxiliary torsion fields ${\cal H}_{\mu \nu \rho}$ and $\mathfrak{C}_{\alpha \beta \gamma}$, 
\begin{equation}\label{I13}
{\cal H}_{\mu \nu \rho}:= \frac{\sqrt{-g}}{\kappa}\,\mathfrak{C}_{\mu \nu \rho}\,, \qquad \mathfrak{C}_{\alpha \beta \gamma} :=C_\alpha\, g_{\beta \gamma} - C_\beta \,g_{\alpha \gamma}+K_{\gamma \alpha \beta}\,.
\end{equation}
Here, $g:=\det(g_{\mu \nu})$, $\sqrt{-g}=\det(e_{\mu}{}^{\hat{\alpha}})$ and $C_\mu$ is the torsion vector $C_\mu :=C^{\alpha}{}_{\mu \alpha} = - C_{\mu}{}^{\alpha}{}_{\alpha}$.
The Einstein field equations can thus be written within the framework of teleparallelism as
\begin{equation}\label{I14}
 \frac{\partial}{\partial x^\nu}\,{\cal H}^{\mu \nu}{}_{\hat{\alpha}}+\frac{\sqrt{-g}}{\kappa}\,\Lambda\,e^\mu{}_{\hat{\alpha}} =\sqrt{-g}\,(T_{\hat{\alpha}}{}^\mu + \mathbb{T}_{\hat{\alpha}}{}^\mu)\,.
\end{equation}
Here, $\mathbb{T}_{\mu \nu}$ is the trace-free energy-momentum tensor of the gravitational field and is given by
\begin{equation}\label{I15}
\kappa\,\mathbb{T}_{\mu \nu} :=C_{\mu \rho \sigma}\, \mathfrak{C}_{\nu}{}^{\rho \sigma}-\frac 14  g_{\mu \nu}\,C_{\rho \sigma \delta}\,\mathfrak{C}^{\rho \sigma \delta}\,.
\end{equation}

The antisymmetry of ${\cal H}^{\mu \nu}{}_{\hat{\alpha}}$ in its first two indices can be used to show that the law of conservation of total energy-momentum tensor in GR$_{||}$, namely,
\begin{equation}\label{I16}
\frac{\partial}{\partial x^\mu}\,\Big[\sqrt{-g}\,(T_{\hat{\alpha}}{}^\mu + \mathbb{T}_{\hat{\alpha}}{}^\mu -\frac{\Lambda}{\kappa}\,e^\mu{}_{\hat{\alpha}})\Big]=0\,,
 \end{equation}
follows from the gravitational field equations. 

Let us recall here that GR field equations can be derived from an action principle involving a gravitational Lagrangian given by
\begin{equation}\label{I17}
L_g = \frac{c ^3}{16 \pi G}\,({^0}R-2\Lambda)\,.
\end{equation}
On the other hand, we find
\begin{equation}\label{I18}
{^0}R=-\frac{1}{2}\,C^{\alpha \beta \gamma}\,\mathfrak{C}_{\alpha \beta \gamma}+\frac{2}{\sqrt{-g}}\,\frac{\partial}{\partial x^\delta}\, \Big(\sqrt{-g}\,C^{\delta}\Big)\,,
\end{equation} 
so that the corresponding Lagrangian for GR$_{||}$ is given by
\begin{equation}\label{I19}
\mathbb{L}_g = - \frac{c ^3}{32 \pi G}\,(C^{\alpha \beta \gamma}\,\mathfrak{C}_{\alpha \beta \gamma}+ 4\,\Lambda)\,.
\end{equation}
The special torsion invariant in Eqs.~\eqref{I18}--\eqref{I19} can be expressed as a linear combination of the three independent algebraic invariants of the torsion tensor, namely, 
\begin{equation}\label{I20}
C^{\alpha \beta \gamma}\,\mathfrak{C}_{\alpha \beta \gamma} = \frac{1}{2}\, C_{\alpha \beta \gamma}C^{\alpha \beta \gamma} + C_{\alpha \beta \gamma}C^{\gamma \beta \alpha} -2\, C_\alpha C^\alpha\,.
\end{equation}

\subsection{GR$_{||}$ as the Gauge Theory of the Translations Group}

Fundamentally, teleparallelism and GR$_{||}$ can only be understood in
the framework of a gauge theory of gravitation~\cite{BlHe}. Nowadays the strong
and the electroweak interactions are described by means of gauge
theories. For gravity this framework can be utilized as well.

Consider first matter in a Minkowski space. The source of gravity in
Newton's theory is the mass density; within special relativity it
should be the energy-momentum tensor $T_{\mu\nu}$ instead. For an
isolated material system, energy-momentum is conserved. This is the
result of the rigid (often called ``global") translation invariance of
the action function of the material system under consideration.

A rigid invariance is in contrast to the idea of field theory. Thus,
in adopting the gauge doctrine, we postulate for the action function
the invariance under {\it local} translations. This forces us to
introduce 1+3 {\it translational gauge potentials} (nonholonomic
frames) $e_\mu{}^{\hat{\a}}$ thereby {\it deforming} the Minkowski
space $M_4$ to a Weitzenb\"ock space $W_4$. Details of this procedure
may be found in Ref.~\cite{Hehl:1979}.

In $W_4$, the Lorentz rotations are not gauged, that is, the
action is still invariant under rigid Lorentz rotations, exactly like
in $M_4$. Accordingly, the $W_4$ connection
$\Gamma_{\mu\hat{\a}}^{\hat{\b}}$ is still flat:
\begin{equation}\label{flat}
  R^{\hat{\a}}{}_{\hat{\b}\mu\nu} :=2\left(  \partial_{[\mu}\Gamma
    _{\nu]\hat{\b}}^{\hat{\a}}   + \Gamma_{[\mu |\hat{\g}} ^{\hat{\a}}
    \Gamma_{|\nu]\hat{\b}}^{\hat{\g}} \right)=0\,.
\end{equation}
This guarantees that in a $W_4$ the parallel transport is still
integrable.  Accordingly, like in $M_4$, we can choose all over
 $W_4$ a {\it suitable} frame such that the connection vanishes
everywhere:
\begin{equation}\label{suitable}
\Gamma_{\mu\hat{\a}}^{\hat{\b}}\stareq 0\qquad\text{(in a suitable
frame)} \,.
\end{equation}

Instead of the curvature, $W_4$ carries a translational field
strength {\it torsion} which, in analogy to electrodynamics, is
represented by the curl of the translational potential
$e_\mu{}^{\hat{\a}}$:
\begin{equation}\label{torsion}
  T_{\mu\nu}{}^{\hat{\a}}:= 2 \nabla_{[\mu} e_{\nu]}{}^{\hat{\a}}
  =2\left(\partial_{[\mu} e_{\nu]}{}^{\hat{\a}}
    + \Gamma^{\hat{\a}}_{[\mu | \hat{\b} |}\,e_{\nu]}{}^{\hat{\b}}\right)\,.
\end{equation}
In the teleparallel frame of Eq.~\eqref{suitable}, we have for the
torsion $T_{\mu\nu}{}^{\hat{\a}}\stareq C_{\mu\nu}{}^{\hat{\a}}\,,$
see Eq.~\eqref{I8}, where $C_{\mu\nu}{}^{\hat{\a}}$ is the object of
anholonomity of Schouten~\cite{Schouten:1954}.
The torsion has three irreducible pieces $^{(I)}{}T_{\mu \nu}{}^\a $, for
$I=1,2,3$. With the torsion vector $T_\mu:= - T_{\mu\nu}{}^\nu$, we have
\begin{equation}\label{irred}
  {}^{(1)}T_{\mu\nu\rho}:= T_{\mu\nu\rho}- {}^{(2)}T_{\mu\nu\rho}
  -{}^{(3)}T_{\mu\nu\rho}\,,\quad {}^{(2)}T_{\mu\nu\rho}
  := - \frac 23 T_{[\mu}\, g_{\nu]\rho}\,,\quad
  {}^{(3)}T_{\mu\nu\rho}:= T_{[\mu\nu\rho]} \,.
\end{equation}

So far we reminded ourselves of the kinematics of a translational
gauge theory (TG). With the gauge Lagrangian
${\cal L}_{\rm TG}={\cal L}_{\rm TG} (\partial e,e,\Gamma,g)$, we can
address the
dynamics by defining the gravitational translational field momentum (or
translation excitation)
\begin{equation}\label{momentum}
  {\cal H}^{\mu\nu}{}_{\hat{\a}}:=-\frac{\partial {\cal L}_{\rm
TG}}{\partial
    T_{\mu\nu}{}^{\hat{\a}}}\,.
\end{equation}
Should we investigate a physical system which has no Lagrangian---in
the case of irreversibility, e.g.---the excitation
$ {\cal H}^{\mu\nu}{}_{\hat{\a}}$ still makes physical sense, as we
know, e.g., from electrodynamics and the inhomogeneous Maxwell
equation.

The general quadratic TG Lagrangian carries three independent pieces:
\begin{equation} \label{telLagr}
{\cal L}_{\text{TG}}\sim \frac 1
  \kappa\left(a_1{}^{(1)}{}T^{\mu\nu}{}_\a\!^{(1)}{}T_{\mu\nu}{}^\a
+a_2{}^{(2)}{}T^{\mu\nu}{}_\a\!^{(2)}{}  T_{\mu\nu}{}^\a
+a_3\,^{(3)}{}T^{\mu\nu}{}_\a\!^{(3)}{}T_{\mu\nu}{}^\a
\right).
\end{equation}
To the Lagrangian \eqref{telLagr} we can add a Lagrange multiplier
term for enforcing the teleparallel constraint, 
see Ref.~\cite{Hehl:1979}. It turns out that {\it we cannot allow spinning
  matter} (other than as test particles) in such a teleparallel space.
Accordingly, we have to decree, see page 52 of Ref.~\cite{Hehl:1979}, that only
scalar and electromagnetic matter be allowed in TG, since
they do not carry dynamical spin and have, as a consequence, 
symmetric energy-momentum tensors.

The translational excitation of Lagrangian \eqref{telLagr} reads,
\begin{equation}\label{momentum}
  {\cal H}^{\mu\nu}{}_\a
=-\frac{\sqrt{-g}}{\kappa}\left(a_1{}^{(1)}{}T^{\mu\nu}{}_\a
    +a_2{}^{(2)}{}  T^{\mu\nu}{}_\a +a_3{}^{(3)}{}T^{\mu\nu}{}_\a \right)\,.
\end{equation}
 In a teleparallelism theory the
three-parametric rigidly Lorentz invariant Lagrangian is a totally
acceptable choice. It corresponds to a gauge theory of the
translation group. However, as it so happens, amongst these
three-parameter Lagrangians, up to an overall constant, there is only
one Lagrangian that is {\it locally Lorentz invariant}, 
see Cho~\cite{Cho}. This theory, which we abbreviate by GR$_{||}$, is, for
scalar and electromagnetic matter, equivalent to GR. The local Lorentz
invariance is imposed from the outside, it is not necessary in a
translational gauge theory. But it shows that GR can be really
understood as a specific translational gauge theory. A
Hilbert-Einstein Lagrangian is equivalent to a definite torsion square
Lagrangian in the teleparallel limit. This is a big step forward in
understanding GR. The constants for GR$_ {||}$ are found to be, see
Ref.~\cite{Muench:1998}:
\begin{equation}\label{GRtel}
a_1=-1\,,\qquad a_2=2\,,\qquad a_3=\frac 12\,.
\end{equation}
This set of constants is called the {\it Einstein choice.} 
Lagrangian~\eqref{telLagr}, together with Eq.~\eqref{GRtel}, and the
attached field momentum~\eqref{momentum} were the starting point for a
classical nonlocal theory of gravity.

\subsection{Nonlocal Gravity}

A locality assumption runs through the standard theories of special and general relativity~\cite{Mash1, Mash2}. For instance, to render an accelerated system in Minkowski spacetime \emph{relativistic}, Lorentz transformations are applied in a pointwise manner all along the world line of the accelerated system. An accelerated observer is thus assumed to be physically identical with a hypothetical inertial observer that shares the same \emph{state}, namely, position and velocity. The locality hypothesis originates from the Newtonian mechanics of classical point particles and its domain of validity is determined by the extent to which physical phenomena could be reduced to pointlike coincidences. However, wave phenomena are generally nonlocal by the Huygens principle. Moreover, Bohr and Rosenfeld have shown that the electromagnetic field measurement requires a certain average over a region of spacetime~\cite{BR1, BR2}. To go beyond the locality assumption, one must include an average over the past world line of the accelerated observer. In this way, a \emph{nonlocal special relativity} theory has been developed~\cite{Mash3a, Mash3b}. 

Can nonlocal special relativity be extended to include the gravitational interaction by means of Einstein's principle of equivalence? Einstein's principle is extremely local, however, and this approach encounters severe difficulties and has been abandoned. Instead, we use Einstein's fundamental insight regarding the connection between inertia and gravitation as a guiding principle and develop nonlocal general relativity patterned after the nonlocal electrodynamics of media. To this end, we exploit the formal analogy between GR$_{||}$ and electrodynamics and introduce an average of the gravitational field into the field equations via a causal constitutive kernel~\cite{HM1, HM2, Mash3c}. In nonlocal gravity, the gravitational field is local, but satisfies partial integro-differential field equations. 

In nonlocal gravity, as in the electrodynamics of media, we retain the gravitational field equations~\eqref{I14}, but change the local constitutive relation~\eqref{I13} to
\begin{equation}\label{N1}
{\cal H}_{\mu \nu \rho} = \frac{\sqrt{-g}}{\kappa}(\mathfrak{C}_{\mu \nu \rho}+ N_{\mu \nu \rho})\,,
\end{equation}
where the new tensor $N_{\mu \nu \rho}$ involves a linear average of the torsion tensor over past events. More specifically, we assume that
\begin{eqnarray}\label{N2}
N_{\mu \nu \rho} = - \int \Omega_{\mu \mu'} \Omega_{\nu \nu'} \Omega_{\rho \rho'}\, {\cal K}(x, x')\,X^{\mu' \nu' \rho'}(x') \sqrt{-g(x')}\, d^4x' \,,
\end{eqnarray} 
where $\Omega(x, x')$ is Synge's \emph{world function}~\cite{Sy}, ${\cal K}$ is the scalar \emph{causal} kernel of the nonlocal theory and  $X_{\mu \nu \rho}(x)$ is a tensor that is antisymmetric in its first two indices and is given by
\begin{equation}\label{N3}
X_{\mu \nu \rho}= \mathfrak{C}_{\mu \nu \rho}+ p\,(\check{C}_\mu\, g_{\nu \rho}-\check{C}_\nu\, g_{\mu \rho})\,.
\end{equation}
Here, $p\ne 0$ is a constant dimensionless parameter and  $\check{C}^\mu$ is the torsion pseudovector defined via the Levi-Civita tensor $E_{\alpha \beta \gamma \delta}$ by
\begin{equation}\label{N4}
\check{C}_\mu :=\frac{1}{3!} C^{\alpha \beta \gamma}\,E_{\alpha \beta \gamma \mu}\,.
\end{equation}

Finally, the gravitational field equation of nonlocal gravity (NLG) is given by
\begin{equation}\label{N5}
 \frac{\partial}{\partial x^\nu}\,\Big[\frac{\sqrt{-g}}{\kappa}\,(\mathfrak{C}^{\mu \nu}{}_{\hat{\alpha}}+N^{\mu \nu}{}_{\hat{\alpha}})\Big]+\frac{\sqrt{-g}}{\kappa}\,\Lambda\,e^\mu{}_{\hat{\alpha}} =\sqrt{-g}\,(T_{\hat{\alpha}}{}^\mu + \mathcal{T}_{\hat{\alpha}}{}^\mu)\,,
\end{equation}
where the energy-momentum tensor of the gravitational field, $\mathcal{T}_{\mu \nu}$, is now given by 
\begin{equation}\label{N6}
\mathcal{T}_{\mu \nu} = \mathbb{T}_{\mu \nu} + \frac{1}{\kappa}\,\left(C_{\mu \rho \sigma} N_{\nu}{}^{\rho \sigma}-\frac 14\, g_{\mu \nu}\,C_{ \delta \rho \sigma}N^{\delta \rho \sigma}\right)\,.
\end{equation}
The total energy-momentum conservation law then takes the form
\begin{equation}\label{N7}
\frac{\partial}{\partial x^\mu}\,\Big[\sqrt{-g}\,(T_{\hat{\alpha}}{}^\mu + \mathcal{T}_{\hat{\alpha}}{}^\mu -\frac{\Lambda}{\kappa}\,e^\mu{}_{\hat{\alpha}})\Big]=0\,.
 \end{equation}
 
No exact nontrivial solution of the nonlocal field equation~\eqref{N5} is known. In this connection, the main source of difficulty appears to be the complicated relation that introduces nonlocality into the theory, namely, Eq.~\eqref{N2}. In a recent paper~\cite{Puetzfeld:2019wwo}, a simpler form of Eq.~\eqref{N2} has been suggested, where the bitensor $\Omega_{\mu \mu'}$  is replaced by the parallel propagator $- g_{\mu \mu'}$. It remains to determine whether this simplification could  help in generating exact nontrivial solutions of NLG. 

The arbitrary tetrad frame we adopted to develop GR$_{||}$ could be any smooth tetrad frame field in spacetime. At each event, any two tetrad frame fields are related by an element of the \emph{local} Lorentz group. This circumstance is in agreement with the invariance of Einstein's GR under the local Lorentz group, since Einstein's theory ultimately depends only upon the metric tensor $g_{\mu \nu}$.  The introduction of nonlocality into the theory may remove this pointwise 6-fold degeneracy of GR$_{||}$. However, as expected, NLG remains invariant under the global Lorentz group.

\section{Nonlocal GEM}

We are interested in the stationary gravitational field of a rotating astronomical body, which is assumed to be confined to a compact region of space. We work in the linear approximation of nonlocal gravity, since the gravitational field is assumed to be weak. In this regime, a certain analogy with classical electrodynamics~\cite{HeOb} turns out to be fruitful. Indeed, in linearized GR, the framework of gravitoelectromagnetism (GEM) has proved rather useful in describing and interpreting the gravitational effects of rotating masses. It is therefore interesting to develop this method in NLG. A preliminary account is already contained in Ref.~\cite{Mash3c} and will be further developed in this paper.

\subsection{Linearized NLG}

In the weak-field regime, we can write the chosen tetrad frame field in the form
\begin{equation}\label{M1}
 e^\mu{}_{\hat{\alpha}}=\d^\mu _{\alpha} -\psi^\mu{}_{\alpha}\,, \qquad  e_\mu{}^{\hat{\alpha}}={\d}_\mu ^{\alpha}+\psi^{\alpha}{}_\mu\,,   
\end{equation}
where $\psi_{\mu \nu}(x)$ is the first-order perturbation away from a background global inertial reference frame in Minkowski spacetime such that   
\begin{equation}\label{M2}
 e^\mu{}_{\hat{\alpha}} \,e_\mu{}^{\hat{\beta}} =\d^{\beta} _{\alpha}.   
\end{equation}
In this linear approximation scheme, the distinction between spacetime and tetrad indices disappears and it follows from Eq.~\eqref{I4} that $g_{\mu \nu} = \eta_{\mu \nu} + \psi_{\mu \nu} + \psi_{\nu \mu}$. Therefore, we can write
\begin{equation}\label{M3}
\psi_{\mu \nu} = \psi_{(\mu \nu)} + \psi_{[\mu \nu]}\,, \quad  g_{\mu \nu} = \eta_{\mu \nu} + h_{\mu \nu}\,, \quad h_{\mu \nu} := 2\,\psi_{(\mu \nu)}\,, \quad \phi_{\mu \nu} := 2\,\psi_{[\mu \nu]}\,.
\end{equation}
The gravitational perturbation $\psi_{\mu \nu}$ is thus comprised of a \emph{symmetric} metric part $\frac{1}{2}\,h_{\mu \nu}$ and an \emph{antisymmetric} tetrad part $\frac{1}{2}\,\phi_{\mu \nu}$. 
In connection with the metric part, it is useful to introduce, as in GR, the trace-reversed potentials
\begin{equation}\label{M4}
\overline{h}_{\mu \nu}=h_{\mu \nu}-\frac 12\eta_{\mu \nu}h\,, \qquad   h:=\eta_{\mu \nu}h^{\mu \nu}\,, \qquad \overline{h}=-h\,.
\end{equation} 
In the teleparallel approach to gravity, the sixteen gravitational potentials consist of ten  metric potentials familiar from GR and six local Lorentz potentials connected with the local choice of the  tetrad system involving three rotations and three boosts. Gravitational potentials are gauge dependent.  Under an infinitesimal coordinate transformation, $x^\mu \mapsto x'^\mu=x^\mu-\epsilon^\mu(x)$, the potentials change to linear order in accordance with $\psi_{\mu \nu} \mapsto \psi'_{\mu \nu}=\psi_{\mu \nu}+\epsilon_{\mu,\nu}$. Hence,
\begin{equation}\label{M5}
\overline{h}\,'_{\mu \nu}=\overline{h}_{\mu \nu}+\epsilon_{\mu,\nu}+\epsilon_{\nu,\mu}-\eta_{\mu \nu}\,\epsilon^\alpha{}_{,\alpha}\,, \quad \overline{h}\,'=\overline{h}-2\,\epsilon^\alpha{}_{,\alpha}\,, \quad   \phi'_{\mu \nu}=\phi_{\mu \nu}+\epsilon_{\mu,\nu}-\epsilon_{\nu,\mu}\,.
\end{equation}

As expected, the linearized gravitational field as well as the corresponding field equations remains invariant under gauge transformations. For instance, it is straightforward to show that the torsion tensor
\begin{equation}\label{M6}
C_{\mu \sigma \nu}=\partial_\mu \psi_{\nu \sigma}-\partial_\sigma \psi_{\nu \mu}
\end{equation}
and the auxiliary torsion tensor  
\begin{equation}\label{M7}
\mathfrak{C}_{\mu \sigma \nu}=-\overline{h}_{\nu [\mu,\sigma]}-\eta_{\nu [\mu}\overline{h}_{\sigma ]\rho,}{}^\rho+\frac 12\phi_{\mu \sigma, \nu}+\eta_{\nu [\mu} \phi_{\sigma ] \rho,}{}^\rho\,
\end{equation}
do not change under a gauge transformation. To obtain the field equations of linearized NLG, we set $\Lambda = 0$ and note that Eq.~\eqref{N5} reduces in the linear regime to
\begin{equation}\label{M8}
\partial_\sigma \,(\mathfrak{C}_{\mu}{}^{\sigma}{}_{\nu} + N_{\mu}{}^{\sigma}{}_{\nu}) = \kappa\, T_{\mu \nu}\,.
\end{equation}
Here, $T_{\mu \nu}$,  $T^{\mu \nu}{}_{,\mu} = 0$,  is the conserved symmetric energy-momentum tensor of matter. We can write 
\begin{equation}\label{M9}
\partial_\sigma\, \mathfrak{C}_{\mu}{}^{\sigma}{}_{\nu}=\, ^{0}G_{\mu \nu} = -\frac
12\,\square\, 
\overline{h}_{\mu \nu}+\overline{h}^\rho{}_{(\mu,\nu)\rho}-\frac
12\eta_{\mu \nu}\overline{h}^{\rho \sigma}{}_{,\rho \sigma}\,,
\end{equation}
where $\square :=\eta^{\alpha \beta}\partial_\alpha \partial_\beta$. Furthermore, 
\begin{equation}\label{M10}
N_{\mu}{}^{\sigma}{}_{\nu} (x) = \int \mathcal{K}(x, y) X_{\mu}{}^{\sigma}{}_{\nu}(y)~d^4y\,,                 
\end{equation}
where $X_{\mu \nu \rho}$ is given by Eq.~\eqref{N3} and $\mathcal{K}(x, y)$ reduces to a universal convolution kernel $K(x-y)$ in the linear approximation~\cite{Mash3c}. The linearized field equations of NLG thus take the GR form
\begin{equation}\label{M11}
 ^{0}G_{\mu \nu} + \partial_\sigma\,N_{\mu}{}^{\sigma}{}_{\nu} = \kappa\,  T_{\mu \nu}\,.              
\end{equation}

For further discussion, see Refs.~\cite{Blome:2010xn, Mash4a, Mash4b} and the references cited therein. 

\subsection{Kernel of Linearized NLG}

A detailed discussion of the causal universal convolution kernel $K(x-y)$ and its reciprocal $R(x-y)$ is contained in Ref.~\cite{Mash3c}. Here, we simply summarize their main properties.

We assume that the convolution kernels $K$ and $R$ are $L^1$ and $L^2$ functions on spacetime. They are reciprocal of each other and satisfy the reciprocity integral equation
\begin{equation}\label{K1}
K(x-y)+R(x-y)+ \int K(x-z)R(z-y)~d^4z = 0\,.
\end{equation}
If $\hat{K}(\xi)$ and $\hat{R}(\xi)$ are Fourier integral transforms of $K(x)$ and $R(x)$, respectively, then, $(1+\hat{K})(1+\hat{R})=1$. Thus, given $\hat{R}(\xi)$, one can in principle determine $K$ from
\begin{equation}\label{K2}
\hat{K} (\xi) =- \frac{\hat{R} (\xi)}{1+\hat{R} (\xi)}\,, 
\end{equation}
 provided $1+\hat{R} (\xi)\ne 0$.
 
We assume that the reciprocal kernel is given by
\begin{equation}\label{K3}
R(x-y)= \nu ~ e^{-\nu\,(x^0-y^0-|\mathbf{x}-\mathbf{y}|)}~\Theta \big(x^0-y^0-|\mathbf{x}-\mathbf{y}|\big)~q(\mathbf{x}-\mathbf{y})\,,
\end{equation}
where $\nu^{-1}$ is a galactic length scale, $\Theta$ is the Heaviside unit step function such that $\Theta(t) = 0$ for $t<0$ and $\Theta(t) = 1$ for $t \ge 0$. Moreover, the Newtonian reciprocal kernel $q(\mathbf{x}-\mathbf{y})$ has been determined on the basis of the observational data regarding the rotation curves of spiral galaxies. Two possible forms for $q$ are given by 
\begin{equation}\label{K4}
q_1=\frac{1}{4\pi \lambda_0}~ \frac{1+\mu_0 (a_0+r)}{r(a_0+r)}~e^{-\mu_0 r}\,
\end{equation}
and
\begin{equation}\label{K5}
q_2=\frac{1}{4\pi \lambda_0}~ \frac{1+\mu_0 (a_0+r)}{(a_0+r)^2}~e^{-\mu_0 r}\,,
\end{equation}
where $r=|\mathbf{x}-\mathbf{y}|$ and $\lambda_0$, $\mu_0$ and $a_0$ are constant parameters such that $\lambda_0$, the fundamental length scale of NLG, is expected to be 
$\approx 3$ kpc and $\mu_0^{-1} \approx 17$ kpc~\cite{RaMa}. The short-distance nonlocality parameter $a_0$ is expected to be much smaller than $\lambda_0$. From the solar system data for the orbit of Saturn, one expects approximately that $a_0$ is  greater than or about the size of the solar system~\cite{Ior, DX, ChMa}. 

Let us now return to Eq.~\eqref{K3} for the reciprocal kernel and note that 
\begin{equation}\label{K6}
\int ~ \nu ~ e^{-\nu\,(x^0-y^0-|\mathbf{x}-\mathbf{y}|)}~\Theta \big(x^0-y^0-|\mathbf{x}-\mathbf{y}|\big)~dy^0=1\,.
\end{equation}
This has an important implication for gravitational fields that are independent of time. For instance, let $Z(y)$ be a smooth function that is independent of $y^0$, then,  
\begin{equation}\label{K7}
 \int R(x-y)\,Z(\mathbf{y})~d^4y= \int q(\mathbf{x}-\mathbf{y})\,Z(\mathbf{y})~d^3y\,.
\end{equation} 
It follows from the reciprocity relation that 
\begin{equation}\label{K8}
 \int K(x-y)\,Z(\mathbf{y})~d^4y= \int \chi(\mathbf{x}-\mathbf{y})\,Z(\mathbf{y})~d^3y\,,
\end{equation} 
where $\chi$ here is the Newtonian kernel reciprocal to $q$; that is, 
\begin{equation}\label{K9}
\chi(\mathbf{x}-\mathbf{y})+q(\mathbf{x}-\mathbf{y})+\int \chi(\mathbf{x}-\mathbf{z})\,q(\mathbf{z}-\mathbf{y})~d^3z=0\,.
\end{equation}

\subsection{GEM in Linearized NLG}

Consider the stationary gravitational field of a rotating astronomical source in the linear approximation. Equation~\eqref{M10} then takes the form
\begin{equation}\label{G1}
N_{\mu}{}^{\sigma}{}_{\nu} (x) = \int K(x-y) X_{\mu}{}^{\sigma}{}_{\nu}(\mathbf{y})~d^4y = \int \chi(\mathbf{x}-\mathbf{y})\,X_{\mu}{}^{\sigma}{}_{\nu}(\mathbf{y})\,d^3y\,.                 
\end{equation}
Next,  from $\partial \chi/\partial x^i =-\partial \chi/\partial y^i$ and Gauss's theorem, we find
\begin{equation}\label{G2}
\partial_\sigma \, N_{\mu}{}^{\sigma}{}_{\nu} = \int \chi(\mathbf{x}-\mathbf{y})\, \partial_i \,X_{\mu}{}^{i}{}_{\nu}(\mathbf{y})\,d^3y\,.
\end{equation}
It proves convenient to impose the gauge conditions
\begin{equation}\label{G3}
\overline{h}^{\mu\nu}{}_{, \nu}=0\,, \qquad \phi_{\mu \nu} = 0\,,
\end{equation}
which correspond to the transverse gauge condition in the metric part and the vanishing of the local Lorentz potentials. The gauge is not quite fixed, however, since a gauge transformation with $\epsilon_\mu = \partial_\mu \zeta$ and $\square\, \zeta = 0$ is still possible. 

With the imposition of  gauge condition~\eqref{G3}, the torsion pseudovector vanishes, $\check{C}^\mu = 0$, by Eqs.~\eqref{N4} and~\eqref{M6}; then, the gravitational field equations of linearized NLG for a stationary source take the form
\begin{equation}\label{G4}
^{0}G_{\mu \nu}(\mathbf{x}) + \int \, \chi(\mathbf{x}-\mathbf{y})\,^{0}G_{\mu \nu}(\mathbf{y})\,d^3y = \kappa\,T_{\mu \nu}\,,  
\end{equation}
since $X_{\mu}{}^{\sigma}{}_{\nu} = \mathfrak{C}_{\mu}{}^{\sigma}{}_{\nu}$ and $\partial_\sigma\, \mathfrak{C}_{\mu}{}^{\sigma}{}_{\nu}=\, ^{0}G_{\mu \nu}$. Equation~\eqref{M9} implies that with the transverse gauge condition, we have $\square\, \overline{h}_{\mu \nu} = - 2\,^{0}G_{\mu \nu}$.  It  then follows from the temporal independence  of gravitational potentials and the reciprocity relation that 
\begin{equation}\label{G5}
\nabla^2\,\overline{h}_{\mu \nu}(\mathbf{x}) = -2\kappa\, [T_{\mu \nu}(\mathbf{x})+\int q(\mathbf{x}-\mathbf{y})\,T_{\mu \nu}(\mathbf{y})~d^3y]\,.
\end{equation}

We assume the source consists of slowly moving matter ($|\mathbf{v}| \ll c$) of density $\rho$, pressure $P$ and matter current ${\bf j}=\rho\, {\bf v}$. The matter energy-momentum tensor can thus be written as $T_{00}=\rho c^2$, $T_{0i}= -c\,j_i$ and $T_{ij}\sim \rho v_iv_j+P\delta_{ij}$. The corresponding gravitational potentials are $\overline{h}_{00}=-4\Phi /c^2$, $\overline{h}_{0i}=-2A_i/c^2$ and $\overline{h}_{ij} = O(c^{-4})$. In the gravitational potentials, we neglect all terms of $O(c^{-4})$. The static gravitoelectric and gravitomagnetic potentials are thus given by 
\begin{equation}\label{G6}
\nabla^2\Phi (\mathbf{x}) = 4\pi G[\rho(\mathbf{x})+\rho_D(\mathbf{x})]\,, \qquad \rho_D(\mathbf{x})=\int q(\mathbf{x}-\mathbf{y}) \rho(\mathbf{y})d^3y\,
\end{equation}
and
\begin{equation}\label{G7}
\nabla^2\,\mathbf{A}(\mathbf{x}) = -\frac{8 \pi G}{c}\, [\,\mathbf{j}(\mathbf{x})+ \mathbf{j}_D(\mathbf{x})]\,, \qquad \mathbf{j}_D(\mathbf{x}) = \int q(\mathbf{x}-\mathbf{y})\,\mathbf{j}(\mathbf{y})~d^3y\,,
\end{equation}
respectively. Here, $\rho_D(\mathbf{x})$ and $\mathbf{j}_D(\mathbf{x})$ are the effective dark matter density  and current, respectively. The transverse gauge condition requires that $\boldsymbol{\nabla} \cdot \mathbf{A}=0$. On the other hand, $\boldsymbol{\nabla} \cdot \mathbf{j} =0$ follows from the energy-momentum conservation law. It follows that the dark matter current is conserved as well; that is, $\boldsymbol{\nabla} \cdot \mathbf{j}_D=0$. 

Let us next introduce the GEM fields $\mathbf {E}_g=\boldsymbol{\nabla} \Phi$ and $\mathbf{B}_g= \boldsymbol{\nabla} \times \mathbf{A}$ such that
\begin{equation}\label{G8} 
\boldsymbol{\nabla} \cdot \mathbf {E}_g=4\pi G\,(\,\rho+\rho_D)\,, \qquad \boldsymbol{\nabla} \times \mathbf{E}_g=0\,,
\end{equation}
\begin{equation}\label{G9}   
\boldsymbol{\nabla} \cdot (\frac{1}{2} \mathbf{B}_g)=0\,, \qquad  \boldsymbol{\nabla} \times (\frac{1}{2}{\bf
B}_g)=\frac{4\pi G}{c}(\, \mathbf{j}+\mathbf{j}_D)\,.
\end{equation}
We remark in passing that our GEM conventions are in conformity with the gravitational Larmor theorem~\cite{Mashhoon:2003ax}.

The corresponding GEM metric takes the form
\begin{equation}\label{G10} 
ds^2=-c^2\left(1+2\frac{\Phi}{c^2}\right)dt^2-\frac{4}{c}({\bf 
A}\cdot d{\bfx})dt+\left(1-2\frac{\Phi}{c^2}\right) \delta_{ij}dx^idx^j\,,
\end{equation}
whose geodesics can be employed to investigate the motion of test particles and null rays in nonlocal GEM. It is possible to show,  for instance, the existence of the gravitational analog of the Lorentz force law~\cite{Mashhoon:2003ax}. We note that $\Phi(\mathbf{x})$ is the gravitoelectric potential of nonlocal gravity in the Newtonian regime and has been investigated in some detail~\cite{Mash3c}; therefore, we concentrate here first on the gravitomagnetic vector potential $\mathbf{A}(\mathbf{x})=O(c^{-1})$.

It is interesting to compare nonlocal GEM with the standard GR treatment~\cite{Mashhoon:2000he, Mashhoon:2003ax}. In NLG, the steady-state assumption is rather necessary and leads to great simplification. Thus topics such as time-varying gravitomagnetism or gravitational induction that are standard in the GR treatment are beyond the reach of nonlocal GEM. Furthermore, 
the steady-state  requirement limits any further gauge freedom; however, it is possible to shift the magnitude of the the gravitoelectric potential by a constant in metric~\eqref{G10}.  To this end, let us recall that the remaining gauge freedom is in the form $\epsilon_\mu = \partial_\mu \zeta$, where $\square\, \zeta = 0$. With $\zeta = - \beta\,(3 t^2 + |\mathbf{x}|^2)/6$, metric~\eqref{G10} changes to 
\begin{equation}\label{G10a} 
ds^2=-\left(1+2\,\Phi + 2\,\beta \right)dt^2- 4\,({\bf 
A}\cdot d{\bfx})\,dt+\left(1-2\,\Phi - \frac{2}{3}\,\beta \right) \delta_{ij}dx^idx^j\,,
\end{equation}
where $\beta$ is a constant parameter.

\section{Gravitomagnetism in Nonlocal GEM}

In Eq.~\eqref{G7}, the gravitomagnetic vector potential depends on the choice of the reciprocal kernel $q$. To indicate which Newtonian reciprocal kernel is under consideration, we introduce a parameter $\delta$ such that $\delta = 1$ for $q_1$ and $\delta = 2$ for $q_2$. Let us write the solution of Eq.~\eqref{G7} in the form
\begin{equation}\label{G11}
{^\delta}\mathbf{A}(\mathbf{x}) = \frac{2\,G}{c}\, \int {^\delta}\mathbb{A}(\mathbf{x} - \mathbf{y})\,\mathbf{j}(\mathbf{y})\,d^3y\,.
\end{equation}
Using Eq.~\eqref{G7} and the explicit form of the Newtonian reciprocal kernels~\eqref{K4}--\eqref{K5}, we find
\begin{equation}\label{G12}
{^\delta}\mathbb{A}(\mathbf{r}) = \frac{1}{r}\, [1 + \alpha_0\,{^\delta}\mathbb{N}(\mu_0\,r)]\,
\end{equation}
where   $r = |\mathbf{r}|$, $\alpha_0 := 2 / (\lambda_0\,\mu_0) \approx 11$ and $^\delta\mathbb{N}$ is given by
\begin{equation}\label{G13}
{^\delta}\mathbb{N}(u) = 1-e^{-u}  + \frac{\delta}{2}\,\varsigma\, e^{\varsigma} [E_1(\varsigma +u)-E_1(\varsigma)] + \frac{1}{2}\,u\, E_1(\varsigma +u)\,.
\end{equation}

Here, $E_1$ is the \emph{exponential integral function} given by
\begin{equation}\label{G14}
E_1(x):=\int_{x}^{\infty}\frac{e^{-t}}{t}dt\,,
\end{equation}
so that for $x: 0 \to \infty$, $E_1(x)$ is a positive monotonically decreasing function that diverges as $-\ln x$ near $x=0$ and falls off exponentially as $x \to \infty$. Moreover, we have introduced a  dimensionless quantity $\varsigma$ such that $\varsigma := a_0 \,\mu_0 <1$. For the exterior of the Earth, we may assume $r$ is small compared to an astronomical unit and therefore $r \ll a_0$, as $a_0$ is about the size of the solar system.  For $r \ll a_0$, we have $\mu_0\,r \ll \varsigma$; then, the Taylor expansion of $E_1(\varsigma +\mu_0\,r)$ about $E_1(\varsigma)$ and repeated differentiation of Eq.~\eqref{G14} result in
\begin{equation}\label{G15}
e^{\varsigma}\,E_1(\varsigma +\mu_0\,r) -  e^{\varsigma}\,E_1(\varsigma) = \frac{r}{a_0} - \frac{1}{2} (1+\varsigma)\, \frac{r^2}{a_0^2} + O\left( \frac{r^3}{a_0^3}\right)\,.
\end{equation}
Putting these results together and neglecting terms of $O\left(r^3 / a_0^3\right)$, we find
\begin{equation}\label{G16}
{^\delta}\mathbb{A}(\mathbf{x} - \mathbf{y}) = \frac{1}{|\mathbf{x} - \mathbf{y}|} + \frac{1}{\lambda_0}\,[2 -\delta + e^{\varsigma}\,E_1(\varsigma)] 
- \frac{1+\varsigma}{2\,\lambda_0\,a_0} (2-\delta) \,|\mathbf{x} - \mathbf{y}|\,.
\end{equation}

Assuming $|\mathbf{x}| > |\mathbf{y}|$, which is appropriate for the exterior of the source, and expanding $|\mathbf{x} - \mathbf{y}|$ to first order in $|\mathbf{y}|/|\mathbf{x}|$, we get
\begin{equation}\label{G17}
\int \frac{\mathbf{j}(\mathbf{y})}{|\mathbf{x} - \mathbf{y}|}\,d^3y \approx \frac{1}{|\mathbf{x}|}\, \int \mathbf{j}(\mathbf{y})\,d^3y +\frac{1}{|\mathbf{x}|^3}\, \int (\mathbf{x} \cdot \mathbf{y})\,\mathbf{j}(\mathbf{y})\,d^3y\,
\end{equation}
and
\begin{equation}\label{G18}
\int |\mathbf{x} - \mathbf{y}| \,\mathbf{j}(\mathbf{y})\,d^3y \approx |\mathbf{x}|\, \int \mathbf{j}(\mathbf{y})\,d^3y -\frac{1}{|\mathbf{x}|}\, \int (\mathbf{x} \cdot \mathbf{y})\,\mathbf{j}(\mathbf{y})\,d^3y\,.
\end{equation}

Let the compact gravitational source reside in the interior of a finite closed spatial domain $\mathcal{D}$ that completely surrounds the source. This means that $\mathbf{j}$ vanishes on the surface of $\mathcal{D}$ and beyond. Then, the conservation of matter current implies
\begin{equation}\label{G19}
\int_{\mathcal{D}} f(\mathbf{y})\, (\mathbf{x} \cdot \mathbf{y})\,\boldsymbol{\nabla}_{\mathbf{y}} \cdot \mathbf{j}(\mathbf{y})\,d^3y = 0\,,
\end{equation}
where $f(\mathbf{y})$ is a smooth function.
Applying Gauss's theorem and setting the integral on $\partial \,\mathcal{D}$ equal to zero, we get
\begin{equation}\label{G20}
\int_{\mathcal{D}} \boldsymbol{\nabla}_{\mathbf{y}}[f(\mathbf{y})\, (\mathbf{x} \cdot \mathbf{y})]\cdot \mathbf{j}(\mathbf{y})\,d^3y = 0\,.
\end{equation}
For $f(\mathbf{y}) = 1$ and $f(\mathbf{y}) = y^i$, we find the following relations
\begin{equation}\label{G21}
\int_{\mathcal{D}} \mathbf{j}(\mathbf{y})\,d^3y = 0\,,\qquad \int_{\mathcal{D}} (\mathbf{x} \cdot \mathbf{y})\,j^i(\mathbf{y})\,d^3y = - \int_{\mathcal{D}} y^i\, \mathbf{x} \cdot \mathbf{j}(\mathbf{y})\,d^3y\,,
\end{equation}
respectively. Let 
\begin{equation}\label{G22}
\int_{\mathcal{D}} \mathbf{y} \times \mathbf{j}(\mathbf{y})\,d^3y = \mathbf{J}\,
\end{equation}
be the total proper \emph{angular momentum} of the gravitational source. Then, it is straightforward to show using Eq.~\eqref{G21} that 
\begin{equation}\label{G23}
 \int_{\mathcal{D}} (\mathbf{x} \cdot \mathbf{y})\,\mathbf{j}(\mathbf{y})\,d^3y = \frac{1}{2}\, \mathbf{J} \times \mathbf{x}\,.
\end{equation}

It then follows from these results that the gravitomagnetic vector potential is given by
\begin{equation}\label{G24}
{^\delta}\mathbf{A}(\mathbf{x}) = \frac{G}{c}\,\frac{\mathbf{J} \times \mathbf{x}}{|\mathbf{x}|^3} \left[ 1 + (2-\delta)\, \frac{|\mathbf{x}|^2}{L_N^2}\right]\,,
\end{equation}
where the relevant nonlocality length scale $L_N$ is given by
\begin{equation}\label{G25}
L_N = \left(\frac{2\,\lambda_0\,a_0}{1+\varsigma}\right)^{1/2}\,.
\end{equation}
The nonlocal contribution to $\mathbf{A}$ at the level of approximation under consideration is nonzero for $q_1$ but \emph{vanishes} for $q_2$. The length scale $L_N \gtrsim 1$ pc, so that the nonlocal contribution to $\mathbf{A}$ in the exterior of the Earth is relatively quite small and less than about $10^{-10}$ of the standard GR value. 

Finally, the gravitomagnetic field can be calculated from Eq.~\eqref{G24} and the result is 
\begin{equation}\label{G26}
{^\delta}\mathbf{B}_g(\mathbf{x}) = \frac{G}{c}\,\frac{3\,(\mathbf{J}\cdot \mathbf{x})\,\mathbf{x} - \mathbf{J}\,|\mathbf{x}|^2}{|\mathbf{x}|^5} + \frac{G}{c}\,\left(\frac{2-\delta}{L_N^2}\right)\,\frac{(\mathbf{J}\cdot \mathbf{x})\,\mathbf{x} + \mathbf{J}\,|\mathbf{x}|^2}{|\mathbf{x}|^3}\,.
\end{equation}
The gravitomagnetic field of the Earth has been directly measured via the GP-B experiment and the GR prediction has been verified to about 19\%~\cite{Francis}. The nonlocal contribution to the gravitomagnetic field of the Earth  is at most ten orders of magnitude smaller than the GR value and is thus beyond current measurement capabilities for the foreseeable future. A similar estimate holds for nonlocal gravitomagnetic effects in the motion of the Moon. In connection with the lunar laser ranging experiment, we note that the main relativistic effects in the motion of the Moon are due to the gravitational field of the Sun and have been calculated in Refs.~\cite{MaTh1, MaTh2}. The Earth-Moon system with its orbital angular momentum acts as an extended gyroscope in the gravitomagnetic field of the Sun. The nonlocal modification of this field is given by Eq.~\eqref{G26} and the corresponding nonlocal gravitomagnetic effects in the motion of the Moon are then about ten orders of magnitude smaller than the GR predictions as well. Another consequence of the existence of the gravitomagnetic field is the Lense-Thirring effect, see 
 Refs.~\cite{MaHeT,ILRC, Renz} and the references cited therein.

\subsection{Nonlocal Contributions to the Metric}

The spacetime metric~\eqref{G10a} in our nonlocal GEM contains gravitoelectric and gravitomagnetic potentials. The latter is given by Eq.~\eqref{G24}. It is therefore necessary to find the corresponding gravitoelectric potential ${^\delta}\Phi (\mathbf{x})$, which is given by
\begin{equation}\label{G27}
{^\delta}\Phi(\mathbf{x}) =  - G\, \int {^\delta}\mathbb{A}(\mathbf{x} - \mathbf{y})\,\rho(\mathbf{y})\,d^3y\,.
\end{equation} 

To simplify matters, we assume that the gravitational source has a spherically symmetric matter distribution. This means that $\rho(\mathbf{y}) = \rho(|\mathbf{y}|)$; then, we go through essentially the same steps as in Eqs.~\eqref{G12}--\eqref{G18}, except that
\begin{equation}\label{G28}
 \int (\mathbf{x} \cdot \mathbf{y})\,\rho(|\mathbf{y}|)\,d^3y = 0\,,
\end{equation}
as a consequence of spherical symmetry for the matter distribution. Therefore, Eq.~\eqref{G16} impies
\begin{equation}\label{G29}
{^\delta}\Phi(r) =- \frac{GM}{r} - \frac{GM}{\lambda_0}\,[2 -\delta + e^{\varsigma}\,E_1(\varsigma)] 
+\frac{GM\,r}{L_N^2}\,(2-\delta)\,
\end{equation}
and
\begin{equation}\label{G30}
{^\delta}\mathbf{E}_g(\mathbf{x}) = \frac{GM\,\mathbf{x}}{|\mathbf{x}|^3}\, \left[ 1 + (2-\delta)\, \frac{|\mathbf{x}|^2}{L_N^2}\right]\,.
\end{equation}
Here, $M$ is the mass of the spherical rotating source in our linear approximation scheme, namely,
\begin{equation}\label{G31}
M := \int  \rho(|\mathbf{y}|) \,d^3y\,.
\end{equation}
We note that Eq.~\eqref{G29} here is consistent with Eqs. (8.39) and (8.40) of Ref.~\cite{Mash3c}. With gravitoelectric potential~\eqref{G29} and gravitomagnetic potential~\eqref{G24}, the GEM metric~\eqref{G10} can now be used consistently in investigating nonlocal effects in GEM.

\subsection{Gravitomagnetic Clock Effect in NLG}

There is a special temporal structure around a rotating mass that is best expressed via the gravitomagnetic clock effect~\cite{CoMa,MGT,MGL,MIL}. To illustrate this effect in NLG, let us assume that the gravitational source rotates about the $z$ axis, $\mathbf{J} = J\, \hat{\mathbf{z}}$, and write the GEM metric in the corresponding spherical polar coordinates. Under the transformation $x^\mu \mapsto (t, r, \theta, \phi)$, metric~\eqref{G10} takes the form
\begin{equation}\label{G32} 
ds^2 =  g_{tt}\,dt^2 + 2\,g_{t\phi} \,dt d\phi + dr^2 + r^2\, d\theta^2 + r^2 \sin^2 \theta \,d\phi^2\,,
\end{equation}
where $g_{tt} := -1- 2\,\Phi$, $g_{t\phi} := - 2\,r\,\sin \theta \, A_{\phi}$ and  we have neglected in our GEM approach the contribution of $\Phi$ to the spatial part of the metric. In the local theory (GR), we have $\Phi = - GM/r$ and $A_{\phi} = (GJ/r^2) \sin \theta$. These potentials change in our nonlocal approach as follows
\begin{equation}\label{G33} 
\Phi = -\mathcal{C}_{\delta} - \frac{GM}{r}\,\left[ 1 - \frac{r^2}{L_N^2} (2-\delta)\right]\,, \qquad  A_{\phi} = \frac{GJ}{r^2}\,\left[ 1 + \frac{r^2}{L_N^2} (2-\delta)\right] \sin \theta\,,
\end{equation}
where
\begin{equation}\label{G34} 
\mathcal{C}_{\delta} = \frac{GM}{\lambda_0}\,[2 -\delta + e^{\varsigma}\,E_1(\varsigma)] \,.
\end{equation}
We are interested in the nonlocal modification of Keplerian periods of the  equatorial circular orbits in this spacetime. 

The geodesic equation for the radial coordinate takes the form
\begin{equation}\label{G35} 
\frac{d^2r}{d\tau^2} + {^0}\Gamma^{r}_{\mu \nu} \,\frac{dx^\mu}{d\tau}\,\frac{dx^\nu}{d\tau} = 0\,,
\end{equation}
where $\tau$ is the proper time. This equation can be solved for $r =$ constant and $\theta = \pi/2$. The solution in the linear approximation under consideration is given by 
\begin{equation}\label{G36} 
\frac{dt}{d\phi} = -\frac{{g_{t\phi}}_{,r}}{{g_{tt}}_{,r}} \pm \left(-\frac{2\,r}{{g_{tt}}_{,r}}\right)^{1/2}\,.
\end{equation}
Indeed, for $\theta = \pi/2$, we have
\begin{equation}\label{G37} 
{g_{tt}}_{,r} = -2\,\Phi_{,r} = -2\,\frac{GM}{r^2}\left[ 1 + \frac{r^2}{L_N^2} (2-\delta)\right]\,, \qquad {g_{t\phi}}_{,r} = \frac{2\,J}{r^2}\left[ 1 - \frac{r^2}{L_N^2} (2-\delta)\right]\,. 
\end{equation}

It follows from a detailed analysis that, as expected, deviations exist from the standard GR results for $\delta = 1, 2$. For an equatorial circular orbit with Keplerian frequency $\omega_K = (GM/r^3)^{1/2}$ and Keplerian period $T_K = 2\pi/\omega_K$, we find for the periods of co-rotating (+) and counter-rotating (-) orbits in terms of coordinate time
\begin{equation}\label{G36} 
t_{\pm}  = T_K \left[1 - \frac{r^2}{2L_N^2}(2-\delta)\right] \pm 2\pi \frac{J}{M}\left[1 - 2\frac{r^2}{L_N^2}(2-\delta)\right]\,  
\end{equation}
and in terms of proper time 
\begin{equation}\label{G37} 
\tau_{\pm}  = T_K \left(1 - \frac{3M}{2r} - \Delta_M\right) \pm 2\pi \frac{J}{M}\left(1 + \frac{3M}{2r} - \Delta_J\right)\,.  
\end{equation}
Here, we work to linear order in perturbation quantities and the nonlocal contributions are given by terms proportional to $r^2/L_N^2$, $\Delta_M $ and $\Delta_J$, where 
\begin{equation}\label{G38} 
\Delta_M  = \mathcal{C}_{\delta} + \frac{r^2}{2L_N^2}(2-\delta)\,,  \qquad  \Delta_J  = \mathcal{C}_{\delta} + 2\,\frac{r^2}{L_N^2}(2-\delta)\,.  
\end{equation}
It is interesting to note that the prograde period is \emph{longer} than the retrograde period, namely,
\begin{equation}\label{G39} 
t_{+} - t_{-}  = 4\pi \frac{J}{M}\left[1 - 2\frac{r^2}{L_N^2}(2-\delta)\right]\,, \qquad \tau_{+} - \tau_{-} = 4\pi \frac{J}{M}\left(1 + \frac{3M}{2r} - \Delta_J\right)\,.  
\end{equation}

In GR, the gravitomagnetic clock effect for circular equatorial orbits around the Earth is given by $\tau_{+} - \tau_{-} \approx 2\times 10^{-7}$ sec. This prediction of GR has not yet been verified by observation.  The GR  effect is indeed rather difficult to measure since the Keplerian period of a near-Earth orbit increases by about $2\times 10^{-7}$ sec when the orbital radius is increased by $0.015$  cm, see Refs.~\cite{Tar1, Tar2, LGM, Ior1, Ior2, ILM, IoLi, LiIoM, Hackmann:2014aga} and the references cited therein. The magnitude of the nonlocal contribution to the gravitomagnetic clock effect for the Earth is smaller than about $10^{-10}$ of the GR value.

\subsection{Gravitational Larmor Theorem in NLG}

In classical electrodynamics, Larmor's theorem establishes a local relation between the motion of a charged test particle in an electromagnetic field and its motion in the absence of the field, but in an accelerated system of reference. The gravitational version of this theorem is essentially Einstein's principle of equivalence expressed within the GEM framework~\cite{Mashhoon:2003ax, Mashhoon:2000he}. It is useful to point out that the theorem extends to \emph{nonlocal} GEM as well.  

Let us imagine an accelerated observer following a world line $\bar{X}^\mu(\tau)$ in Minkowski spacetime. Here, $\tau$ is the observer's proper time. The observer carries an orthonormal tetrad frame $\lambda^{\mu}{}_{\hat \alpha}(\tau)$ along its path such that 
\begin{equation}\label{L1}
\frac{d\,\lambda^{\mu}{}_{\hat \alpha}}{d\tau}\, = \Psi_{\hat \alpha \hat \beta}\,\lambda^{\mu \hat \beta}\,,
\end{equation}
where $\Psi_{\hat \alpha \hat \beta} = - \Psi_{\hat \beta \hat \alpha}$ is the antisymmetric acceleration tensor. In analogy with the electromagnetic field tensor, we can decompose $\Psi_{\hat \alpha \hat \beta}$ into its ``electric" and ``magnetic" parts, namely, $\Psi_{\hat 0 \hat i} = \gamma_{\hat i}$ and $\Psi_{\hat i \hat j} = \epsilon_{\hat i \hat j \hat k}\,\omega^{\hat k}$. 
Here,  $\boldsymbol{\gamma}$ and $\boldsymbol{\omega}$ represent the invariant translational and rotational accelerations of the observer, respectively. Let us now introduce a geodesic coordinate system in the neighborhood of the accelerated observer. At a given proper time $\tau$, the straight spacelike geodesics normal to $\bar{X}^\mu(\tau)$ form a Euclidean hyperplane. An event on this hyperplane with inertial coordinates $X^\mu$ will be assigned geodesic (Fermi) coordinates $x^{\hat \mu}$ such that $x^{\hat 0} = \tau$ and $X^\mu - \bar{X}^\mu(\tau) = x^{\hat i}\,\lambda^{\mu}{}_{\hat i}(\tau)$. With these transformations, $dS^2 = \eta_{\mu \nu}\,dX^\mu \,dX^\nu$ becomes $dS^2 = g_{\hat \mu \hat \nu}\,dx^{\hat \mu} \,dx^{\hat \nu}$, where 
\begin{align}\label{L2} 
g_{\hat 0 \hat 0}&=-(1+\boldsymbol{\gamma}\cdot \mathbf{x})^2+(\boldsymbol{\omega}\times \mathbf{x})^2\,,\\
\label{L3} g_{\hat 0 \hat i}&=(\boldsymbol{\omega}\times \mathbf{x})_{\hat i}\,,\qquad g_{\hat i \hat j} = \delta_{\hat i \hat j}.
\end{align}
A detailed discussion of these local coordinates and their admissibility is contained in Ref.~\cite{Mash3c}. In general, $\boldsymbol{\gamma}$ and $\boldsymbol{\omega}$ are functions of proper time $\tau$. However, in the present context of steady-state GEM, we assume that these accelerations are constants and do not vary with proper time. 

A comparison of this flat metric at the linear order with metric~\eqref{G10a} once we neglect its spatial curvature reveals that an accelerated observer in Minkowski spacetime is \emph{locally} equivalent to an observer in a GEM field provided  $\Phi + \beta = \boldsymbol{\gamma}\cdot \mathbf{x}$ for a suitable choice of the constant $\beta$ and $-2\,\mathbf{A} = \boldsymbol{\omega}\times \mathbf{x}$, which means that $\mathbf{E}_g = \boldsymbol{\gamma}$ and 
$\mathbf{B}_g = -\boldsymbol{\omega}$, respectively. These GEM fields contain nonlocal effects; in this way, the gravitational Larmor theorem has been extended to the nonlocal regime. 

An interesting application of the gravitational Larmor theorem involves the interaction of spin with the gravitational field. The coupling of intrinsic spin with the gravitomagnetic field has been discussed extensively and a brief review of the subject is contained in Ref.~\cite{Mashhoon:2003ax}. The effect is related to spin-rotation coupling via the gravitational Larmor theorem. The spin-rotation coupling for neutrons has recently been measured via neutron interferometry~\cite{DDSH, DDKWLSH}. The extension of the gravitational Larmor theorem to the nonlocal regime means that spin-gravity coupling can likewise be extended to the nonlocal regime.

\section{Gravitational Energy-Momentum Tensor}

The traceless gravitational energy-momentum tensor $\mathcal{T}_{\mu \nu}$ of NLG is given by Eq.~\eqref{N6}.  Let us first compute the local part of this tensor $\mathbb{T}_{\mu \nu}$, which is traceless as well, for the GEM case. To this end, we write Eq.~\eqref{I15} in the form
\begin{equation}\label{T1}
\kappa\,\mathbb{T}_{\mu \nu} := C_{\mu \rho \sigma}\, \mathfrak{C}_{\nu}{}^{\rho \sigma}-\frac 14 g_{\mu \nu}\,\mathbb{I}\,, \qquad \mathbb{I} := C_{\rho \sigma \delta}\,\mathfrak{C}^{\rho \sigma \delta}\,,
\end{equation}
and express the components of the torsion tensor in terms of the GEM potentials. That is, 
\begin{equation}\label{T2}
c^2 C_{0 i 0} = \Phi_{,i}\,, \quad c^2 C_{0 i j} = A_{j,i}\,, \quad c^2 C_{i j 0} = A_{i,j} - A_{j,i}\,, \quad c^2 C_{i j k} = \delta_{ik}\,\Phi_{,j} - \delta_{jk}\,\Phi_{,i}\,
\end{equation}
and
\begin{equation}\label{T3}
c^2 \mathfrak{C}_{0 i 0} = 2\,\Phi_{,i}\,, \quad c^2 \mathfrak{C}_{0 i j} = A_{j,i}\,, \quad c^2 \mathfrak{C}_{i j 0} = A_{i,j} - A_{j,i}\,, \quad c^2 \mathfrak{C}_{i j k} = O(c^{-2})\,.
\end{equation}
It follows that 
\begin{equation}\label{T4}
\mathbb{I} = \frac{1}{c^4} \left[ 4\,E_g^2 -3\,B_g^2 -2\,\sum_{i,j}\,A_{(i,j)}\,A_{(i,j)}\right]\,,
\end{equation}
where we have used the relation
\begin{equation}\label{T5}
\sum_{i,j} A_{i,j}\,A_{i,j} = \frac{1}{2}\, B_g^2 + \sum_{i,j}A_{(i,j)}\,A_{(i,j)}\,.
\end{equation}

It is now possible to compute the components of the traceless energy-momentum tensor, which are
\begin{equation}\label{T6}
\mathbb{T}_{0 0} := - \frac{1}{8 \pi G} \left[E_g^2 + \frac{1}{4}\,B_g^2 -\frac{1}{2}\,\sum_{i,j}\,A_{(i,j)}\,A_{(i,j)}\right]\,,
\end{equation}
\begin{equation}\label{T7}
\mathbb{T}_{0 i} := \frac{1}{8 \pi G}\, (\mathbf{E}_g \times \mathbf{B}_g)_i\,
\end{equation}
and
\begin{equation}\label{T8}
\mathbb{T}^{ij} :=  \frac{1}{8 \pi G} \left[2\, (E^i_g\,E^j_g -  \frac{1}{2}\, \delta_{ij} E_g^2) + (B^i_g\,B^j_g - \frac{1}{2}\,\delta_{ij} B_g^2)- \mathcal{A}^{ij}\right]\,,
\end{equation}
where
\begin{equation}\label{T9}
\mathcal{A}_{ij} :=  \sum_{k}\,A_{k,i}\,A_{k,j}- \frac{1}{2}\,\delta_{ij}\,\sum_{m,n}\,A_{m,n}\,A_{m,n}\,.
\end{equation}
These local results must be supplemented with nonlocal terms. That is, we must go back to Eq.~\eqref{N6} and compute $\mathcal{T}_{\mu \nu}$, which contains nonlocal terms of the form
\begin{equation}\label{T9a}
N_{\mu \nu \rho}(\mathbf{x})= \int \chi(\mathbf{x}- \mathbf{y})\,\mathfrak{C}_{\mu \nu \rho}(\mathbf{y})\,d^3y\,,
\end{equation}
where $\chi$ is the kernel  of NLG theory in the Newtonian regime~\cite{Mash3c}.  The explicit calculation of this kernel is rather complicated and is beyond the scope of this paper.

It is interesting to compare and contrast the local Eqs.~\eqref{T6}--\eqref{T9} with those obtained via the Landau-Lifshitz pseudotensor $t_{\mu\nu}$ of GR~\cite{L+L} within the standard GEM framework~\cite{Mashhoon:2003ax, Mashhoon:2000he}. To this end, it is necessary to assume a steady-state GR configuration (i.e., $\partial \Phi /\partial t=0$ and
$\partial \mathbf{A}/\partial t=0$). Then, 
\begin{align}\label{T10} 
4\pi G\,t_{00}&=-\frac{7}{2}E_g^2+\sum_{i,j}A_{(i,j)} A_{(i,j)}\,,\\
\label{T11} 4\pi G\,t_{0i}&=2\,(\mathbf{E}_g\times \mathbf{B}_g)_i\,,\\
\label{T12} 4\pi G\,t^{ij}&=\left(E_g^iE_g^j-\frac{1}{2}\delta_{ij}E_g^2\right)+\left( 
B_g^iB_g^j+\frac{1}{2}\delta_{ij}B_g^2\right)\,.
\end{align}

The similarity between these different gravitational results and the corresponding electromagnetic ones is noteworthy. In particular, imagine a steady-state configuration involving a rotating astronomical source with mass $M$ and angular momentum $\mathbf{J } = J\, \hat{\mathbf{z}}$. Then, it follows from the gravitational Poynting vector that there is a steady circulation of gravitational energy in the same sense as the rotation of the source with an azimuthal flow speed given in spherical polar coordinates by 
\begin{equation}\label{T13} 
v_g \propto \frac{J}{Mr}\sin \theta\,.
\end{equation}
The proportionality constant depends on the underlying  theory of gravitation~\cite{Mashhoon:2003ax, Mashhoon:2000he}.

\section{Discussion}

We have developed gravitoelectromagnetism (GEM) within the framework of nonlocal gravity (NLG). Except for the trivial solution of field equations involving flat spacetime, NLG has no other known exact solution at present. We must therefore resort to the linearized theory, where GEM is possible for steady-state configurations. We have examined the nonlocal GEM corrections to the stationary gravitational field of an isolated rotating mass in the weak-field and slow-motion approximations. Due to the existence of galactic length scales in NLG, the nonlocal GEM effects around the Earth or the Sun turn out to be at most about ten orders of magnitude smaller than the corresponding GR effects.   \\   \\





\end{document}